\documentclass[a4paper,12pt]{article}
\usepackage{graphicx}
\begin{document}
\newcommand{\tanb}{\mbox{$\tan \! \beta$}}
\newcommand{\mer}{m_{{\tilde{e}_R}}}
\newcommand{\mlsp}{m_{{\tilde{\chi}^0_1}}}
\newcommand{\sig}{\sigma_{\chi p}}
\newcommand{\mwi}{m_{\tilde{\chi}^+_1}}
\newcommand{\lsp}{\tilde{\chi}^0_1}
\newcommand{\gsim}{\buildrel>\over{_\sim}}
\newcommand{\lsim}{\buildrel<\over{_\sim}}

\setcounter{page}{0}
\thispagestyle{empty}
\begin{flushright}
YITP--01--33 \\
April 2001 \\
\end{flushright}

\vspace{2cm}

\begin{center}
{\Large \bf Implications of Muon anomalous magnetic moment
for Direct detection of Neutralino Dark Matter}

\baselineskip=32pt

Yeong Gyun Kim and Mihoko M. Nojiri

\baselineskip=22pt

{ YITP, Kyoto University, Kyoto, 606-8502, Japan} 

\end{center}

\vspace{1cm}

\begin{abstract}
\noindent
We investigate the implications of the recent measurement of muon
anomalous magnetic moment for the direct detection of neutralino dark
matter in the three different SUSY models: mSUGRA, a model with
non-universal Higgs mass, and an $SO(10)$ GUT model. We consider two
extreme scenario for $\Delta a_\mu$ bound, i.e.  $27 \times 10^{-10} <
\Delta a_\mu < 59 \times 10^{-10}$ (1$\sigma$ bound) and 
$0 < \Delta a_\mu < 11 \times 10^{-10}$ ($2\sigma$ below).  
In mSUGRA model, the
counting ratio may be above the sensitivity of the future experiments
when parameters are within $1\sigma$ bound of $\Delta a_{\mu}$.
However, the $\Omega_{\chi}$ tends to be high compared to the currently
favored value $\Omega=0.3$.
For models with the non-universal scalar masses, the possibility to have
the consistent $\Omega_{\chi}$ and the high counting ratio is open up in the
region of parameter space where Higgsino mass $\mu$ is smaller than
mSUGRA prediction. In particular, in the $SO(10)$ model, 
the LSP dark matter detection rate may be enhanced by almost
one order of magnitude compared to mSUGRA and the model with
non-universal Higgs mass, for cosmologically acceptable
$\Omega_{\chi} h^2$.  The highest detection rate of LSP dark matter
occurs in the region where the LSP constitutes a subdominant part of
local halo DM.
Implication of SUSY mass parameter measurement under the cosmological
constraint is also discussed.
\end{abstract}

\vspace{2cm}

\vfill

\pagebreak

\baselineskip=14pt

\section{Introduction}

The Minimal Supersymmetric Standard Model (MSSM) \cite{SUSY} is one of
the best motivated extensions of the Standard Model. It offers a
natural solution of the hierarchy problem \cite{witten} as well as
amazing gauge coupling unification \cite{amaldi}. Since naturalness
requires that at least some superparticles have masses at or below the
TeV scale, supersymmetric theories generally predict a rich
phenomenology at future colliders such as Tevatron, LHC, or 
proposed linear colliders (LC) \cite{colrev}. As an extra ``bonus'',
the simplest version of the MSSM, where $R-$parity is conserved, also
contains a new stable particle (the lightest supersymmetric particle,
LSP); in most cases this is the lightest neutralino, which often makes
a good Dark Matter(DM) candidate \cite{jungman}.

Though we might have to wait for LHC experiment to start 
for the direct discovery of supersymmetry, 
we may be able to  probe SUSY models earlier 
by using precision measurements 
in low energy experiments.
Sparticles contribute to low energy measurements 
through loop effects, and these effects may become significant if
the their masses are not too large. Another possibility 
is direct detection of dark matter by the detector placed
deep underground \cite{DAMA,CDMS1,CDMS2,cresst,genius}.
Note that the existence of dark matter already indicates
the needs of new physics.

Recently the Brookhaven E821 experiment has released a measurement of
the muon anomalous magnetic moment, reporting a $2.6 \sigma$
deviation from the standard model value \cite{BNL}
\begin{eqnarray} \label{data}
\Delta a_{\mu} \equiv a_{\mu}^{exp} - a_{\mu}^{SM} = 43(16) \times 10^{-10},
\end{eqnarray} 
which has generated considerable interest.
This deviation may indeed be a sign of new physics beyond the standard
model (SM) and could be accommodated by supersymmetric contributions 
\cite{cza}--\cite{baek}.
There are essentially two types of diagrams involving
superparticles which contribute to $a_\mu$, i.e., neutralino--smuon
and chargino--sneutrino loop diagrams \cite{amugen}.  Since $a_\mu$ requires
chirality violation, for $\tanb \gg 1$ the dominant contributions are
proportional to the product of an electroweak gauge coupling and the
Yukawa coupling of the muon, where the latter factor either comes
directly from the higgsino component of the chargino or neutralino in
the loop, or from $\tilde{\mu}_L - \tilde{\mu}_R$ mixing. For $\tanb
\gg 1$ one has thus $\Delta a_\mu \propto \tanb$.  If mass splittings
between different sparticles are not too large, so that their mass
scale can be described by the single parameter $m_{\rm SUSY}$, the
overall result can be estimated as \cite{moroi}
\begin{eqnarray} \label{amu}
\left| \Delta a_\mu \right| = {1 \over 32 \pi^2} \left( \frac{5}{6}
g_2^2 + \frac{1}{6} g_1^2 \right)
{m_\mu^2 \over m_{\rm SUSY}^2} \tan\beta
\end{eqnarray}
where $g_1$ and $g_2$ are the $U(1)_Y$ and $SU(2)$ gauge couplings,
respectively. In our convention,
the sign of $\Delta a_\mu$ is equal to the sign of 
Higgsino mass parameters $\mu$,
when the gaugino masses and \tanb\ are positive. 
Therefore $\Delta a_{\mu}$ imposes non-trivial constraints on 
the MSSM parameters such as the sign of $\mu$,
superparticle masses and the ratio \tanb\ of 
the vacuum expectation values (vevs) of the two neutral
Higgs fields of the MSSM.
The new data prefers a positive $\mu$ and one has an upper or lower 
limit on the chargino
and sneutrino masses for a given $\tanb$.
Through these non-trivial constraints, $\Delta a_{\mu}$ will affect
the predictions of other observables.

There exist relations between the SUSY contribution to $a_\mu$
and the predicted LSP--nucleon scattering cross section \cite{drees}. 
A direct connection comes from the common
dependence on \tanb, since both $a_\mu$ and $\sig$ increase with
increasing \tanb. 
Similar to the SUSY contribution to $a_\mu$,
the leading contributions to the
spin--independent (coherent) contribution to $\sig$ also involve
violation of chirality \cite{dn5}.
Main contributions come from the
exchange of CP--even Higgs bosons. 
It couples to the strangeness component of proton or neutron, 
whose coupling  has the same \tanb\
dependence as the Yukawa coupling of the muon. 
Note also that the sign of $\mu$, favored by $a_{\mu}$ measurement, 
suggests  large value of $\sig$ for smaller LSP mass 
due to the lack of accidental cancellation of 
couplings \cite{drees}.
Importance of DM search experiments have now increased substantially.  


In general, larger $\sig$ implies smaller relic DM density $\Omega_\chi h^2$.  
Since the Higgs--LSP--LSP
couplings require higgsino--gaugino mixing, they scale like $1/\mu$
for $\mu^2 \gg M_Z^2$. The Higgs exchange contribution to $\sig$ then
scales like $1/\mu^2$. 
On the other hand, the change in the size of $\mu$ 
also affects $\Omega_\chi h^2$
through the LSP dark matter annihilation into gauge bosons
or s--channel Higgs boson.
The annihilation cross section for $\chi\chi \rightarrow WW$ and $ZZ$
are highly suppressed in the bino--like LSP case which is favored in 
mSUGRA model.
However, a large cross section is possible when the LSP has a significant 
Higgsino component.
This is because $WW$ cross section depends on $W \chi \chi^\pm$ interactions
and only couplings $W \tilde H^0 \tilde H^\pm$ and 
$W \tilde W^0 \tilde W^\pm$
are allowed by gauge invariance. 
For $ZZ$ cross section, it depends on $Z \chi \chi_i^0$ interactions which
is possible only through $Z \tilde H^0 \tilde H^0$ couplings.
Hence a reduction of $\mu$ (an increase in higgsino component of the LSP) 
implies an increase of LSP annihilation cross section, which in turn
gives a reduction of $\Omega_\chi h^2$. 
The pair annihilation into fermions through s--channel pseudoscalar
Higgs boson is also enhanced if LSP is the mixture of gaugino and higgsino.

The relations between $a_{\mu}$ and $\sig$, and 
$\sig$ and $\Omega_\chi h^2$ we discussed so far are 
model independent. One may further investigates more model 
dependent aspect of the relation. This has been intensively studied in 
a class of model called mSUGRA with an universal scalar mass 
$m$, an universal gaugino mass $M$, 
and an universal trilinear coupling $A$ \cite{drees,ellis,arnowitt,baek}. 
The model relates a lower bound of higgs mass and $\Delta a_{\mu}$
to the constraints to right handed slepton and LSP masses, 
and a simple flavor structure 
of soft masses leads the prediction to $b\rightarrow s \gamma$. 

The parameter space of mSUGRA is then quite limited. 
This is because the model predicts $\mu\gg M_1$ unless 
$\tan\beta\gg 1$ or $m\gg M$ . 
This tends to lead to too large a $\Omega_{\chi} h^2$, because 
the LSP pair annihilation cross section,$\sig(\chi\chi\rightarrow X)$ is
proportional to $m_\chi^2 /m_{\tilde l_R}^4$ while 
Higgs mass and $a_{\mu}$ constraints give a lower bound of 
those particles masses in general. 
The $b\rightarrow s \gamma$ constraint 
also pushes up the overall SUSY scale. 
Especially for large $\tanb~ (\sim 30)$, one might need stau-neutralino
coannihilation or a s--channel annihilation of the neutralino 
to make relic DM density reasonable \cite{largetb}.
The DM signal also turns out to be rather low; the maximal 
value of the cross section is only slightly above the 
proposed sensitivities. 

Note however that the constraint strongly relies on the 
assumption that all scalar masses are universal at GUT scale. 
The rather parameter independent prediction $\mu\gg M$ 
in mSUGRA is actually relying on the strict universality.  
Smaller value of $\mu$ can be easily 
achieved in a model with non-universal Higgs masses \cite{drees}. 
This leads to consistent DM density and larger counting ratios. 
The constraint from $b\rightarrow s\gamma$ is sensitive to 
on the structure of scalar masses. 

A study of non-mSUGRA model may be useful in this stage 
where the experimental constraints start to strongly limit the parameter 
space of mSUGRA. Given the upper limit $\sig$
in mSUGRA, discovery of DM in the early stage of CDMS II experiment 
would enforce us to face to more flexible models . 
In this paper, we investigate the implications of recent measurement
of muon anomalous magnetic moment 
for direct detection of Neutralino Dark Matter \cite{sigold,dn5,signew}
in the various SUSY models.
We impose the accelerator bounds of Higgs and sparticle masses.
However, the prediction for $b \rightarrow s\gamma$ decay rate
is sensitive to details of the flavor
structure of the soft breaking terms, unlike the quantities we 
consider here. We therefore do not attempt to analyze the
constraint from $b \rightarrow s \gamma$ decays quantitatively
\footnote{In mSUGRA, the constraint from $b \rightarrow s\gamma$ decay
rate is not important compared to that from Higgs mass for $\mu > 0$ case 
\cite{ellis2}.}.

The muon anomalous magnetic moment will be further checked 
using four times more data collected already.
On the other hand, it can be consistent with 
the standard model value if one takes into account 
other theoretical calculations of 
the hadronic vacuum polarization \cite{yn}.
In this regards, we consider two extreme cases for the range of $\Delta a_\mu$;
\begin{eqnarray} 
27 \times 10^{-10} < \Delta a_{\mu} < 59 \times 10^{-10}  
~~~~~  (1\sigma~\it{bound}),
\end{eqnarray} 
and
\begin{eqnarray}
0 < \Delta a_{\mu} < 11 \times 10^{-10}  
~~~~~  (2\sigma~\it{below})
\end{eqnarray} 

We study the minimal supergravity (mSUGRA) model as
well as some other SUSY models where the assumption of strict scalar
mass universality at the GUT scale is relaxed.  
Unlike mSUGRA, the relic density of neutralino could be 
very small for those models. 
Since the direct DM detection rate depends on the product of the $\sig$
and local LSP density $\rho_\chi$, one must consider what fraction
of our local halo density $\rho_{local}$ could be composed of neutralinos.
Unless there is separation for different types of DM,
the ratio of LSP DM to total DM should be the same locally in the Galaxy
and globally in the whole Universe. In our paper we present the thermal 
relic density 
$\Omega_{\chi}$ and $\sig$ separately. However, following assumption 
might be useful to estimate the signal ratio,
\begin{eqnarray}
\rho_\chi = \rho_{local} \times ({\Omega_\chi \over \Omega_{DM}})
\end{eqnarray}
where $\Omega_{DM}$ is the total contribution of DM to the total
energy density of the Universe.
When combined with the reduced Hubble constant $h \simeq 0.7$ and 
$\Omega_{DM} \simeq 0.3$ we have $\Omega_{DM} h^2 \simeq 0.15$.
For our present study, we assume that $\rho_\chi = \rho_{local}$
if $\Omega_\chi h^2 \geq 0.15$, and present the contours of 
future experimental reach using local LSP density rescaled as 
$\rho_\chi = \rho_{local} \times (\Omega_\chi h^2 /0.15)$
if $\Omega_\chi h^2 < 0.15$. 
For convenience, we define the scaled cross section, 
\begin{eqnarray}
\sigma_{scaled} &\equiv& \sig \times (\Omega_\chi h^2 / 0.15)
~~\rm{for}~ \Omega_\chi h^2 < 0.15  \nonumber\\
&\equiv& \sigma_{\chi p}~~~~~~~~~~~~~\,~~~~~~~~ 
\rm{for}~ \Omega_\chi h^2 \geq 0.15.
\end{eqnarray}
We will find that the increase in the $\sig$ may compensate 
the decrease in the $\Omega_\chi h^2$
so that the DM detection rate more or less remains unchanged
\footnote{The model independent situation 
is recently discussed in \cite{gondolo}.}.
For some model we also see that more than one order of magnitude enhancement of 
the counting ratio is possible 
when LSP is the subdominant component of the local halo. 

\section{mSUGRA}

In the minimal supergravity model, it is usually assumed that all
squared scalar masses receive a common soft SUSY breaking contribution
$m^2_Q = m^2_U = m^2_D = m^2_L = m^2_E = m^2_{H_u} = m^2_{H_d} \equiv
m^2$ at the GUT scale $M_X \simeq 2 \cdot 10^{16}$ GeV, while all
gauginos receive a common mass $M$ and all trilinear soft terms unify
to $A$ at the same scale. 
The renormalization group (RG) evolution of soft breaking
squared Higgs masses then leads to consistent breaking of the
electroweak symmetry, provided the higgsino mass parameter $\mu$ can
be tuned independently \cite{radbreak}. In this paper, we choose the
weak scale input parameters $m_b(m_b)=4.2$ GeV, $m_t(m_t)=165$ GeV,
and $\tan\beta$. We minimize the tree level potential at
renormalization scale $Q= \sqrt{m_{\tilde{t}}m_t}$, which essentially
reproduces the correct value of $\mu$ obtained by minimizing the full
1--loop effective potential \cite{dn2}.
With these assumptions, the mSUGRA model allows four continuous free
parameters ($m, M, A$ and $\tan\beta$). 
We take $\mu > 0$ because of BNL constraint. 

\begin{figure}[htbp]
\begin{center}
\includegraphics[width=8.5cm,angle=0]{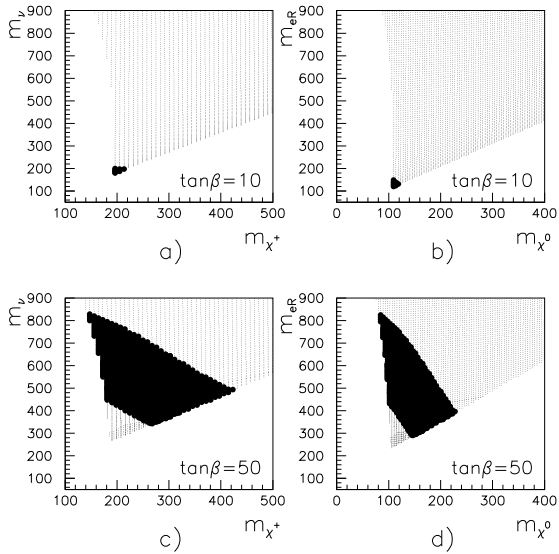}
\end{center}
\caption{\footnotesize Allowed mass ranges of (a) chargino-sneutrino 
and (b) LSP-SU(2) singlet slepton in mSUGRA for $\tanb = 10$. 
$\tanb = 50$ case is shown in (c) and (d).
We take $A = 0$
and $\mu > 0$, and scan $m \in [0,1]$ TeV and $M \in [0, 1]$ TeV,
subject to experimental constraints. The heavily marked points satisfy
1$\sigma$ bound of $\Delta a_\mu$.}
\label{fig1}
\end{figure}

For generic SUSY parameters, the chargino-sneutrino diagram provides 
a dominant contribution to $a_\mu$.
In Fig. 1 (a) and (c), 
we plot the allowed mass range of chargino and sneutrino 
for two different choices of $\tanb = 10$ and 50 respectively. 
The allowed mass range of LSP and SU(2) singlet sleptons are shown
in Fig. 1 (b) and (d) for $\tanb=10$ and 50 respectively.
Here, we take $A=0$ and $\mu > 0$ and allow $m$ and $M$ to
vary in the intervals $m < 1$ TeV and $M < 1$ TeV. 
For the parameter range scanned in this plot, we find that the lightest
Higgs boson $h$ couples essentially like the single Higgs boson of the
SM; we thus demand that its mass $m_h >$ 111 GeV. This follows from
recent LEP results \cite{lephiggs}, allowing for a 2 GeV theoretical
uncertainty in the calculation of $m_h$ \cite{2loop}. 
Further uncertainty on $m_h$ comes from the error of top quark mass value
($m_h$ is varied by $\sim\pm 3$ GeV when $m_t$ is varied by $\pm 5$ GeV around 
175 GeV \cite{ellis3}).
We include loop corrections to the masses of neutral Higgs bosons 
from the third generation quarks and squarks, 
including leading two--loop corrections \cite{2loop}. 
We further require that the chargino mass $\mwi >$ 100 GeV \cite{alnew}. 
We exclude regions where the LSP is charged, i.e., $m_{\tilde
\tau_1} < m_{\tilde{\chi}_1^0}$ or $m_{\tilde{t}_1} < m_{\tilde{\chi}_1^0}$. 
The parameter region is cut by Higgs mass constraint at left edge
of the allowed region, and the lower edge of the allowed region is 
determined by neutral LSP constraint.
The heavily marked points satisfy the further
requirement of $27 \times 10^{-10} < \Delta a_\mu < 59 \times 10^{-10}$. 
We use the expressions of ref.\cite{moroi} for the
calculation of $\Delta a_\mu$. 

For $\tanb = 10$, only small range of sparticle masses
($m_{\chi^\pm} \sim 200$ GeV and $m_{\tilde\nu} \sim 200$ GeV) are allowed by 
the lightest Higgs mass bound, the requirement of the neutral LSP
and 1$\sigma$ bound of $\Delta a_\mu$.
For the small $\tanb$ values, the lower bound of $\Delta a_\mu$ 
gives a strict upper limit on sparticle masses because $\Delta a_\mu$
is proportional to $\tanb$ as mentioned already.
On the other hand, the lightest Higgs
mass bound $m_h > 111$ GeV 
requires lower limit of sparticle masses because large radiative
corrections from top and stop loops are required. 
These two opposite tendency allow only limited region of parameter space
for the small $\tanb$ values.

Upper limit of sparticle masses increase as $\tanb$ increases.
For $\tanb = 50$, the maximal values of light chargino  
and sneutrino mass are $\sim 430$ GeV and $\sim 820$ GeV respectively
within 1$\sigma$ bound of $\Delta a_\mu$.
For large $\tanb$ values, the region of parameter space where
$m_{\chi^\pm}$, $m_{\tilde\nu}$ is small
is excluded by the upper bound of $\Delta a_\mu$ measurement
because small sparticle masses gives 
too large $\Delta a_\mu$ values in this case.

These upper (and lower) bounds of sparticle masses with the fact that
$\mu > 0$ is preferred 
has a significant importance on the predictions 
of neutralino-proton cross section,
$\sig$ \cite{drees,arnowitt} and neutralino dark matter 
relic density, $\Omega_\chi h^2$.
In most situations the dominant contribution to the spin independent
amplitude is the exchange of the two neutral CP--even Higgs bosons.
The sizes of the $\lsp \lsp (h,H)$ couplings
do depend quite significantly on the sign of $\mu$, unless $\tanb \gg
1$ \cite{dn3}. In particular, for $\mu < 0$ strong cancellations occur
\cite{dn5} both within different contributions to the same coupling, and between
the $h$ and $H$ exchange contributions to $\sig$.
Because positive $\mu$ is preferred by $\Delta a_\mu$ experiment,
cancellations would not occur.
Also the upper (lower) limit of sparticle masses has a significant effect on
the minimum (maximum) values of neutralino-proton cross section 
and neutralino dark matter density. 
\begin{figure}[htbp]
\begin{center}
\includegraphics[width=8.5cm,angle=0]{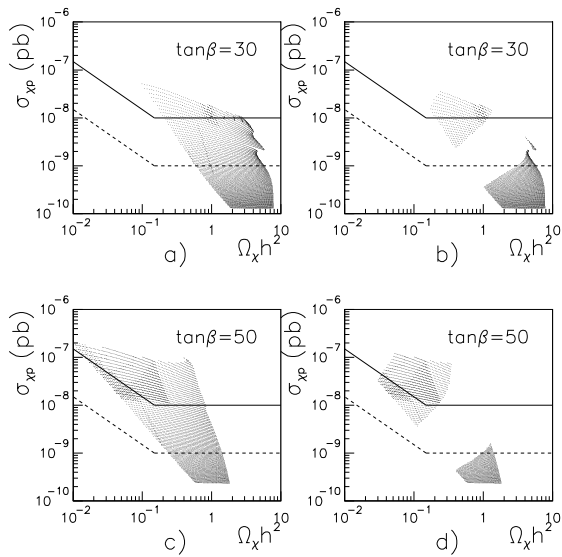}
\end{center}
\caption{\footnotesize $\sig$ vs. $\Omega_\chi h^2$ in mSUGRA
(a) without and (b) with 1$\sigma$ bound of $\Delta a_\mu$ 
for $\tanb = 30$ respectively. 
The corresponding plots are shown in (c) and (d)
for $\tanb = 50$ respectively. We take $A = 0$
and $\mu > 0$, and scan $m \in [0,1]$ TeV and $M \in [0, 1]$ TeV,
subject to experimental constraints.
The straight and dashed lines show 
$\sigma_{scaled}=10^{-8}$ and $10^{-9}$ pb,
the approximate level of 
the claimed sensitivity by the CDMS(Soudan) and GENIUS proposals respectively.
We also indicate the region of parameter space where $\Delta a_\mu$ 
is more than $2\sigma$ lower than current central value
((b) and (d) lower region).}
\label{fig2}
\end{figure}
When the 1 $\sigma$ bound of $\Delta a_{\mu}$ is imposed, 
there is not too much parameter space remaining for $\tan\beta\sim 10$ .
In Fig. 2(a)((c)) and (b)((d)), we therefore show the allowed region 
of $\sig$ vs. $\Omega_\chi h^2$ without and with $\Delta a_\mu$
constraints for $\tanb = 30(50)$ respectively.
Here, we take $A=0$ and $\mu > 0$ and allow $m$ and $M$ to
vary in the intervals $m < 1$ TeV and $M < 1$ TeV
and require $m_h > 111$ GeV, $m_{\chi^\pm} > 100$ GeV and neutral LSP. 
The straight and dashed lines show the approximate level of 
the claimed sensitivity by the CDMS(Soudan)\cite{CDMS2}
and GENIUS \cite{genius} proposal respectively.
The calculation of $\sig$ is based on
refs. \cite{dn5,dd}. We use the value $m_s \langle p | \bar{s} s | p
\rangle = 130$~MeV for the strange quark's contribution to the nucleon
mass; this matrix element is uncertain to about a factor of 2, leading
to a similar uncertainty in the prediction of $\sig$. Finally, the
calculation of the scaled LSP relic density $\Omega_\chi h^2$ uses
results of refs.\cite{dn3,bdd}; $s-$channel poles are treated as
described in ref.\cite{dy}.
The co--annihilation of $\lsp$ with sleptons is not included. 

For $\tanb = 30$, $\sig$ is larger than $\sim 3 \times 10^{-9}$ pb
and can reach upto $\sim 4 \times 10^{-8}$ pb 
within 1$\sigma$ bound of $\Delta a_\mu$ (upper region in Fig. 2 (b)). 
In this case, $\Omega_\chi h^2$ vary from $\sim 0.2$ to $\sim 1.6$.
If we require a cosmological constraint $\Omega_\chi h^2 < 0.3$ \cite{ombound},
some portion of region are allowed and are within 
the approximate level of the claimed sensitivity by CDMS(Soundan)
and the GENIUS proposal. Note that we do not include 
the $\tilde\tau\chi^0_1$ coannihilation here. 
The lower limit of $\Omega_\chi h^2$ might be 
reduced substantially for the points of most left-down
portion of parameter space 
in Fig. 2 (a) and (b) where parameter space is ended due to 
the condition that $\chi^0_1$ is not LSP. 
Note however that the prediction is very sensitive to the running 
of the third generation sparticle $\tilde\tau$
between Planck scale and GUT scale
\cite{barhall, nft}. 
For $\tanb = 50$, $\sig$ is larger than $\sim 4 \times 10^{-9}$ pb
and can reach up to $\sim 1 \times 10^{-7}$ pb 
within 1$\sigma$ bound of $\Delta a_\mu$. 
$\Omega_\chi h^2$ is mostly less than 0.3 due to the effect of 
s-channel pseudoscalar Higgs pole \cite{dn3,largetb,ellis}
of neutralino pair annihilation and is thus
cosmologically acceptable. The smaller value of $m_A$ is achived 
by the bottom Yukawa running of $m_{H_d}$ enhanced by $\tan^2\beta$. 
Heavy higgs exchange contribution in $\sig$ is also enhanced 
as $H\bar{s} s$ coupling is enhanced and $m_H$ is reduced. 
Again all region is within the approximate level of the claimed sensitivity 
by the GENIUS proposal . 

On the other hand, when $a_{\mu}$ turns out to be more than $2\sigma$ 
below the current central value,
$\sig$ is mostly below $1 \times 10^{-9}$ pb and $\Omega_\chi h^2$
is larger than 1(0.4) for $\tan\beta=30(50)$. For $\tan\beta=10$,  
$\sig<0.5 \times 10^{-8}$ pb and $\Omega_\chi h^2>0.4$ 
is found in our numerical study. 
They are  cosmologically disfavored
unless $\tilde{\tau}\tilde{\chi}^0_1$ coannihilation is taken into account.


\section{More general models}

Now we relax our assumptions, allowing for non--universal
soft scalar masses at the GUT scale, while keeping the unification of
the gaugino masses. As specific models, we consider a model with 
non-universal Higgs mass and an $SO(10)$ Grand Unified model \cite{drees}.

In a model with non-universal Higgs, we assume that soft breaking Higgs mass
terms are different from the other universal soft scalar mass terms
at GUT scale,
\begin{eqnarray} \label{nuh}
m \neq m_{H_u} (= m_{H_d})
\end{eqnarray}
For simplicity we keep $m_{H_u} = m_{H_d}$ at the GUT scale
\footnote{A model with non-universal Higgs mass, eq. (\ref{nuh})
can be considered
as a special case of $SO(10)$ model, eq. (\ref{so10}) with $M_D^2 = 0$.}.

The $SO(10)$ theory incorporates a complete generation of MSSM matter
superfields into the 16--dimensional spinor representation,
$\Psi_{16}$. In addition to these matter superfield, the minimal
$SO(10)$ model includes a 10--dimensional Higgs superfield $\Phi_{10}$
which contains the two Higgs superfields of the MSSM (as well as their
$SU(3)$ triplet, $SU(2)$ singlet partners). When $SO(10)$ breaks to
the MSSM gauge group $SU(3)_C \times SU(2)_L \times U(1)_Y$,
additional $D-$term contributions (parameterized by $M_D^2$ which can
be either positive or negative) to the soft SUSY breaking masses arise
\cite{dterm}:
\begin{eqnarray} \label{so10}
m^2_Q = m^2_E = m^2_U = m^2_{16} + M^2_D \nonumber \\
m^2_D = m^2_L = m^2_{16} - 3 M^2_D, ~ 
m^2_{H_{u,d}} = m_{10}^2 \mp 2 M^2_D,
\end{eqnarray}
where $m_{16}$ and $m_{10}$ are scalar soft breaking masses for fields
in the {\bf 16} and {\bf 10} dimensional representations of
$SO(10)$, respectively. We consider this model too, but
we assume $m_{16}=m_{10} \equiv m$ for simplicity.

The modifications (\ref{nuh}) and (\ref{so10}) of the mSUGRA
boundary conditions change our predictions 
through the changed value of
$|\mu|$ and pseudoscalar Higgs mass $m_A$ at the weak scale, 
which mostly affects $\sig$ and $\Omega_\chi h^2$
(and also through modifications of the slepton spectrum, 
which change $\Delta a_\mu$ and $\Omega_\chi h^2$, for a $SO(10)$ model).
The changes of $|\mu|$ value can be understood as follows. In the mSUGRA
scenario, the contributions of $m$ and $M$ to the weak scale values of
the soft breaking Higgs boson masses can be parameterized as
\begin{eqnarray} \label{msugra}
m^2_{H_d} \simeq m^2 + 0.5 M^2; \nonumber \\
m^2_{H_u} \simeq \epsilon_H m^2 - 3.3 M^2 ,
\end{eqnarray}
where we have assumed $\sin\beta \simeq 1$ but ignored contributions
from the bottom Yukawa coupling. The coefficient $\epsilon_H$ is
small, because the GUT scale value of $m^2_{H_u}$ is canceled by the
scalar masses appearing in the RG running \footnote{The effective
$\epsilon_H$ is slightly negative for low SUSY breaking scale, but
turns positive if this scale is large. Recall that the relevant scale
for the analysis of gauge symmetry breaking increases with increasing
sparticle masses.}. The effect of non--universality on $m_{H_u}^2$ can
be parameterized by introducing $m_Q^2 + m_U^2 + m_{H_u}^2 =3 m_s^2$
and $\delta m_H^2 = m_{H_u}^2 - m_s^2$:
\begin{equation} \label{nonuniv}
m_{H_u}^2 = \delta m_H^2 + \epsilon_H m_s^2 - 3.3 M^2.
\end{equation}
This equation follows because the radiative correction to $m^2_{H_u}$
is proportional to $m^2_Q + m^2_U + m^2_{H_u}$, hence the effect of
RGE running is the same as in mSUGRA with the replacement $m^2
\rightarrow m_s^2$.  If $\sin \beta \simeq 1$, correct gauge symmetry
breaking requires $\mu^2 \simeq -m^2_{H_u} -M_Z^2/2$, where all
quantities are taken at the weak scale. 
In mSUGRA ($i.e., \delta m_H^2 = 0$), $\mu > M$ is predicted
unless $m$ is extremely large. However, if $\delta m_H^2$ is non-zero, it
directly affects the value of $|\mu|$ at the weak scale. 

These modifications of the mSUGRA boundary conditions also have an impact on
the Higgs masses. At tree level, the mass of the pseudoscalar Higgs boson
is simply given by 
\begin{eqnarray}
m_A^2 &=& m_{H_u}^2 + m_{H_d}^2 + 2\mu^2 \nonumber \\
     &\simeq& m_{H_d}^2 + \mu^2 - M_Z^2 / 2
\end{eqnarray}
where all quantities are taken at the weak scale and in the second line,
we used the relation $\mu^2 \simeq -m^2_{H_u} -M_Z^2/2$ which holds 
if $\sin \beta \simeq 1$. Therefore, $m_A^2$ depends on $m_{H_d}^2$ and
$\mu^2$ at weak scale. When we increase the soft Higgs masses
in same amounts, the increase of $m_{H_d}^2$ is compensated by the
decrease of $\mu^2$ (through the increase of $m_{H_u}^2$).
In the model with non-universal Higgs mass 
(with the assumption $m_{H_u}^2 = m_{H_d}^2$ at GUT scale),
$m_A^2$ thus does not change too much along the change of $m_H/m$. 
On the other hand, in the $SO(10)$ model, the modifications of 
$m_{H_u}^2$ and $m_{H_d}^2$ at GUT scale are in the opposite direction.
In this model, $m_A^2$ would be quite reduced by the decrease of $m_{H_d}^2$
and $\mu$ when $M_D^2$ has a large and negative value.

The reduction of $m_A$ implies 
the reduction of charged Higgs mass $m_{H^\pm}$ and
heaviest CP even Higgs mass $m_H$. This has a significant implication on
$\Omega_\chi h^2$, $\sig$  
and the branching ratio of $b \rightarrow s\gamma$ decay.
In mSUGRA model, LSP pair annihilation through 
$\chi\chi \rightarrow A,H \rightarrow X$ channel is not important
unless $\tan\beta \gg 1$. However, when $m_A$ and $m_H$ decrease,
the above channel can be quite enhanced especially in the resonance region.
This in turn gives quite reduced $\Omega_\chi h^2$.
On the other hand, the neutralino-proton cross section $\sig$ increases
as $m_H$ decreases.
This is because the dominant contributions to $\sig$ comes from 
the exchange of the two CP-even Higgs bosons.
Also $BR(b \rightarrow s\gamma)$
is affected by $m_H^\pm$ value
through charged Higgs-top quark loop. 
The reduction of $m_H^\pm$ tends to increase 
$BR(b \rightarrow s\gamma)$ because the charged Higgs-top 
quark loop always interferes with SM contribution constructively.
Therefore the preferred value of $\tan\beta$ is expected to 
be different from that of mSUGRA. 
Note that the ino-squark loop have a 
opposite sign to charged Higgs-top loop in mSUGRA model for $\mu > 0$ case. 
For very large $\tan\beta$, 
they tend to overshoot the charged Higgs contribution. 

\begin{figure}[htbp]
\begin{center}
\includegraphics[width=8.5cm,angle=0]{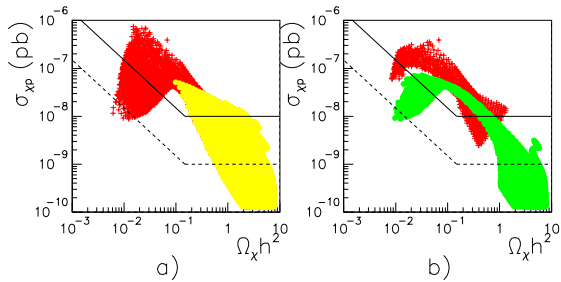}
\end{center}
\caption{\footnotesize $\sig$ vs. $\Omega_\chi h^2$ in a model
with non-unversal Higgs mass
(a) without and (b) with $\Delta a_\mu$ constraint
for $\tanb = 30$ respectively. 
We take $A = 0$ and $\mu > 0$, and scan $m \in [0,1]$ TeV ,
$M \in [0, 1]$ TeV and $1 \leq m_{H_u}^2 / m^2 \leq 20$,
subject to experimental constraints.
The straight and dashed lines show the approximate level of 
the claimed sensitivity by the CDMS(Soudan) and GENIUS proposal respectively.}
\label{fig3}
\end{figure}

Fig.~3 shows the correlation between $\sig$ and $\Omega_\chi h^2$ for
$\tanb = 30$, $A=0$ and $\mu > 0$ in a model with non-universal Higgs mass.
Here, we scan the region $m < 1$ TeV, $M < 1$ TeV and 
$1 \leq m_{H_u}^2 / m^2 \leq 20$, 
under the same mass constraint in Fig.~2.
In Fig.~3(a), the light (yellow) shaded region indicates 
$m_{H_u}^2 / m^2 = 1$, i.e, mSUGRA case. 
The maximum value of $\sig$ increases by about one order, compared to 
mSUGRA case and reaches up to $\sim 7 \times 10^{-7}$ pb where
$\Omega_\chi h^2$ value is about 0.02 and therefore LSP constitutes a
subdominant part of local halo DM.
The highest value of the scaled cross section, $\sigma_{scaled}$ 
is $\sim 10^{-7}$ pb.
In Fig.~3(b), 
we further require that the $\Delta a_\mu$ constraint is satisfied.
In this figure, the dark (red) shaded regions are correspond to
the 1$\sigma$ bound case and the medium (green) shaded regions to
the more than $2\sigma$ below current central value case.
For the 1$\sigma$ bound case, $\Omega_\chi h^2$ values in large portion 
of allowed regions are less than 0.3 and most of such regions are within
the approximate level of the claimed sensitivity by the CDMS(Soudan) proposal.
Also, for the $2\sigma$ below case, the large portion of allowed region
are cosmologically acceptable and within the sensitivity level by the GENIUS,
contrary to mSUGRA model where all regions are cosmologically disfavored
and have very small $\sig$ with $2\sigma$ below case. 

\begin{figure}[htbp]
\begin{center}
\includegraphics[width=11.5cm,angle=0]{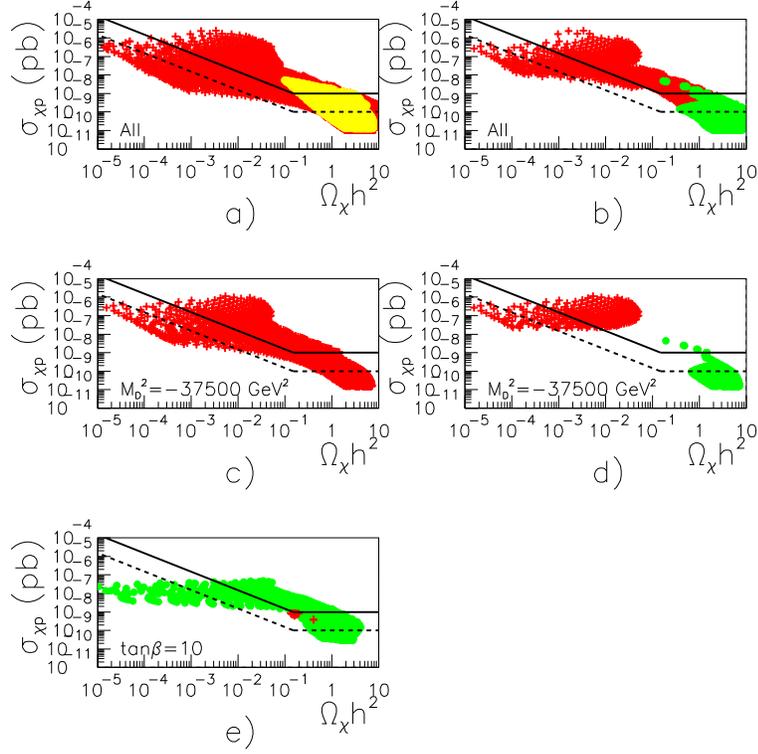}
\end{center}
\caption{\footnotesize $\sig$ vs. $\Omega_\chi h^2$ in 
$SO(10)$ model
(a) without and (b) with $\Delta a_\mu$ constraint
for $\tanb = 30$ respectively. 
We take $A = 0$ and $\mu > 0$, and scan $m \in [0,1]$ TeV ,
$M \in [0, 1]$ TeV and $-50000 \rm{GeV}^2 \leq M_D^2 \leq +50000 \rm{GeV}^2$,
subject to experimental constraints.
The corresponding plots are shown in (c) and (d) 
for fixed $M_D^2=-37500 \rm{GeV}^2$.
We also show the scatter plot in (e) with $\Delta a_\mu$ 
constraint for $\tanb = 10$.
The straight and dashed lines show the approximate level of 
the claimed sensitivity by the CDMS(Soudan) and GENIUS proposal respectively.
}
\label{fig4}
\end{figure}
Fig.~4 shows the correlation between $\sig$ and $\Omega_\chi h^2$ for
$\tanb = 30$, $A=0$ and  $\mu > 0$ in $SO(10)$ model.
Here, we scan the region $m < 1$ TeV , $M < 1$ TeV and 
$-50000 \rm{GeV}^2 \leq M_D^2 \leq +50000 \rm{GeV}^2$, 
under the same mass constraint in Fig.~2
\footnote{We keep all soft scalar masses are positive at GUT scale,
therefore $-{m^2 \over 2} < M_D^2 < {m^2 \over 3}$ always.}.
In Fig.~4(a), the light (yellow) shaded region indicates 
$M_D^2 = 0$, i.e, mSUGRA case.  
The maximum value of $\sig$ increases by more than two order of
magnitude compared to mSUGRA case. It reaches up to $\sim 2 \times 10^{-5}$ pb
where $\Omega_\chi h^2$ value is $\sim 4\times 10^{-3}$.
The maximum value of $\sigma_{scaled}$ is $\sim 2 \times 10^{-6}$ pb
\footnote{This is the approximate level of claimed signal by DAMA\cite{DAMA}.} 
when $\Omega_\chi h^2 \sim 2 \times 10^{-2}$.
Therefore the LSP constitutes a subdominant part of the local halo DM
in this case.
In Fig.~4(b), we further require that $\Delta a_\mu$ bound is satisfied.
In this figure, the dark (red) shaded regions are correspond to
the 1$\sigma$ bound case. 
Large portion of the parameter space have cosmologically allowed 
$\Omega_\chi h^2<0.3$ and within the approximate level of 
the claimed sensitivity by the GENIUS proposal.

For $SO(10)$ model, the Higgs sector is changed drastically, 
as it can be read off directly from eq. (\ref{so10}).
The heavy CP even and CP odd Higgs masses 
can be quite small if $M_D^2$ has large negative values.
In this case, $\sig$ are largely enhanced and the LSP sometimes
hit the Higgs pole leading to greatly reduced $\Omega_\chi h^2$.
There appear the region where the detection rate (which
depends on the product $\sig~ \Omega_\chi$) are enhanced by more than 
one order of magnitude, compared to mSUGRA and non-universal Higgs mass model,
within cosmologically acceptable region. This happens when
$m_A \ll 2 m_\chi$, therefore pole effect is reduced while
$\sig$ is enhanced. 

Such highest detection rate comes from the points where LSP is
a subdominant component of the halo Dark Matter, 
as can be seen in the figure. 
For clarity, we show the correlation between $\sig$ and $\Omega_\chi h^2$
in Fig.~4 (c) without and (d) with $\Delta a_\mu$ constraints 
for $M_D^2 = -37500$ GeV$^2$.  

The medium (green) shaded regions correspond to the parameter 
space where $a_{\mu}$ is lower more than $2\sigma$ compared to 
the current central value. Here the effect of the existence of the D term is
small; $\sig$ remains small and $\Omega_\chi h^2$ is too high. This is 
due to our artificial cut off of the parameter space of $M_D^2$. 
To make $\mu$ small to have acceptable  $\Omega_\chi h^2$, one needs $M_D$ 
of roughly same order to that of gaugino masses. 
To compare we show the scatter plot for $\tan\beta = 10$ in Fig. 4 (e). 
The SUSY scale is substantially smaller compared to in Fig. 4 (b)
($\tan\beta=30$), therefore effect of $M_D^2$ is visible. 
The $2\sigma$ below case 
could have acceptable $\Omega_\chi h^2$ and a large counting ratio
for any value of $\tanb$ when $M_D^2$ is changed freely.  

\section{Conclusions}

In this paper, we have investigated the implications of the recent 
measurement of muon MDM on the spin--independent 
neutralino--proton cross section $\sig$ and LSP dark matter 
relic density $\Omega_\chi h^2$
in several SUSY models with either universal or non--universal
soft scalar masses at the GUT scale. 
We considered two extreme scenario for $\Delta a_\mu$ bound,i.e,
$27 \times 10^{-10} < \Delta a_\mu < 59 \times 10^{-10}$ (1$\sigma$ bound)
and $0 < \Delta a_\mu < 11 \times 10^{-10}$ ($2\sigma$ below).
$\sig$ and $\Delta a_\mu$ can become
large if \tanb\ is large, because
both quantities are sensitive to chirality
violation in the matter (s)fermion sector (which in the MSSM is
enhanced for large \tanb) and if the SUSY mass scale is low.
On the other hand, $\Omega_\chi h^2$ become large if the
SUSY mass scale is high.
The 1$\sigma$ bound of $\Delta a_\mu$ 
prefer high $\tanb$ and low SUSY mass scale
while for the $2\sigma$ below case, 
low $\tanb$ and high SUSY mass scale are preferred.
Therefore 1$\sigma$ bound of $\Delta a_\mu$ gives generally
large $\sig$ and small $\Omega_\chi h^2$ while the opposite
is true for $2\sigma$ below.
While these general statements are fairly model--independent as
long as one sticks to the field content of the MSSM, quantitative
predictions do depend significantly on details of the spectrum of
superparticles, in particular on the implementation of radiative gauge
symmetry breaking.

We considered three different models: mSUGRA, a model with 
non-universal Higgs mass, and an $SO(10)$ GUT model. 
In mSUGRA model, some portion of allowed parameter space is
cosmologically acceptable. The $\sig$ is  
within the sensitivity level by the GENIUS proposal,
if we require $1\sigma$ bound of $\Delta a_\mu$.
On the other hand, if $\Delta a_\mu$ is  $2\sigma$ below the current 
central value, 
the maximum $\sig$ is mostly below $10^{-9}$ pb and 
$\Omega_\chi h^2$ so large that it is cosmologically disfavored
unless coannihilation of $\tilde\tau\chi^0_1$ is taken into account. 

In a model with non-universal Higgs mass 
(for $1 < m_{H_u}^2 / m^2$ and $\tanb = 30$), 
the maximum value of $\sig$ increases by about
one order of magnitude while the minimum value of $\Omega_\chi h^2$
is reduce by about one order of magnitude. 
With the 1$\sigma$ bound of $\Delta a_\mu$, large portion of 
allowed parameter space gives cosmologically acceptable $\Omega_\chi h^2$
and are within the sensitivity level of CDMS(Soudan) proposal.
Even with the case where $\Delta a_\mu$ is 
$2\sigma$ below from current central value, some portion of
parameter space gives cosmologically acceptable $\Omega_\chi h^2$
and are mostly within the sensitivity level of the GENIUS proposal.
In this model, the maximal allowed value of dark matter detection rate
(which depends on the product $\sig~\Omega_\chi$) remains more or less same
as that in mSUGRA model. 

Dramatic deviations from mSUGRA predictions are possible if
one introduces $SO(10)$ $D-$term contributions to scalar masses. In
this model, within the 1$\sigma$ bound of $\Delta a_\mu$, 
the maximal allowed value of $\sig$ for given \tanb\ can
exceed the mSUGRA prediction by almost three order of magnitude, 
if the $D-$term contribution to the mass of the Higgs boson that couples to
the top quark is positive and large. 
The corresponding  $\Omega_\chi h^2$
is also reduced, however the counting ratio maybe enhanced by 
factor of 10 compared to the previous models. 
In this model, the higgs sector is changed drastically. The heavy CP
even and CP odd Higgs masses can be quite small if $M_D^2$ has large
negative values. The counting ratio is the highest when the LSP constitutes
a subdominant potion of local halo DM. 

The predictions of $\sig$ and $\Omega_\chi h^2$ depend on the 
boundary conditions of model parameters at GUT scale. 
The FCNC constraints only require universality in squark sector 
in the same gauge multiplets. The universality of sfermions and higgs
masses at the GUT scale as in mSUGRA requires $\mu\gg M$, and it is not 
favored cosmologically. In mSUGRA model, only the parameter space with 
mass degeneracy of $\tilde{\tau}$ and $\chi^0_1$ or the 
$2m_{\chi}\sim m_A$ is allowed.

The too large relic density  may imply that LSP 
might have more higgsino component than predicted in the mSUGRA 
model, or equivalently non-universal scalar masses at the GUT 
scale for higgs sectors. The improved measurement of $\Delta a_{\mu}$, 
(non) discovery of dark matter at future search experiments, 
and improved observations of $\Omega_\chi h^2$  will give us a hint  on the 
LSP natures. Especially, if DM is discovered in the 
early stage of near future experiments, we should consider 
the variation of the soft mass terms.

The implication to the collider physics is also important. 
Consistency to the cosmology force us to think about the 
parameter space where $\tan\beta\sim 50$ or $m_{\chi^0_1}
\sim$ $m_{\tilde{\tau}}$ in mSUGRA. For example, the degeneracy of stau and LSP 
mass indicates the domination of the decay mode into 
stau and very low acceptance of tau and anti-tau lepton pairs 
in $\chi^0_2\rightarrow  \tilde{\tau}\tau$ decays at LHC. 
The end point of $m_{\tau\tau}$ in the cascade decay is important
for the determination of sparticle masses \cite{colrev}.
However, it is equivalently possible 
that the nature's choice is a non-universal Higgs mass at GUT 
scale. In such case  $\tilde{\chi}^0_4$ and $\tilde{\chi}^+_2$ may have 
substantial wino component and therefore they will be frequently 
produced in $\tilde{q}_L$ decay at LHC. The identification of 
such decay modes and the determination of $\mu$ parameter is 
demonstrated in \cite{dkntk} for the case where SUSY scale 
is relatively low and the cascade decay of 
ino to slepton has substantial branching ratios. 
When SUSY scale is high, the ino with substantial 
higgino and gaugino mixing dominantly 
decays into Higgs bosons. The study for such a 
case might be interesting to see the ability 
for LHC to measure the relation between $\mu$ and $M$
to determine $\sig$ and $\Omega_\chi h^2$.

\subsection*{Acknowledgments}
We thank to S. Jhingan for careful reading of the manuscript.
M.M.N. was supported in part by a Grant-in-Aid 
for Scientific Research from the Ministry of Education (12047217).

\end{document}